\documentclass{osa-article}
\usepackage{caption}
\usepackage{subcaption}
\journal{osac}

\articletype{Research Article}
\begin{document}

\title{Nature of the orbital angular momentum (OAM) fields in a  multilayered fiber}

\author{Ramesh Bhandari}
\address{Laboratory for Physical Sciences, 8050 Greenmead Drive, College Park,  Maryland 20740, USA}

\email{rbhandari@lps.umd.edu} 



\begin{abstract}
We provide a theoretical analysis of the nature of  the  orbital angular momentum (OAM)  modal fields  in a multilayered fiber, such as the step-index fiber and the ring-core fiber. In a detailed study of the vector field solutions of the step-index fiber (in the exponential basis) we discover that  the polarization-induced field component is a modified scalar OAM field (as opposed to a standard OAM scalar field) with a shifted intensity pattern in the weakly guiding approximation (WGA); the familiar intensity donut pattern is reduced or increased in radius depending upon whether it is a case of spin-alignment or anti-alignment with the OAM.  Such a shift in the intensity pattern appears to be  a general feature of the  field of a multilayered fiber as seen from an extension to the  ring-core fiber.
Additionally, we derive a general expression for the polarization-correction to the scalar propagation constant, which includes, for the first time, the contribution of the polarization-induced field.  All the analytic expressions are illustrated and validated  numerically with  application to a step-index fiber, whose analytic solutions are well-known. 
\end{abstract}
\vspace{-0.5mm}
\vspace{-6.0mm}
\section{Introduction}
\vspace{-1.5mm}
Because of the possibility of a multifold increase in data-carrying capacity on an optical fiber, research in orbital angular momentum propagation (OAM) mode propagation in a fiber has proliferated. The orthogonality of the OAM modes makes possible the augmentation of fiber capacity by stacking different streams of data on different OAM modes, characterized by a unique set of  topological charge, radial index number,  and polarization,  but possessing the same wavelength. The topological charge, usually denoted $l$, is representative of  the orbital angular momentum $l\hbar$ per photon carried by the mode. 
\\\\
The    vector     OAM modes are the solutions,  $HE_{l+1,m}$ and $EH_{l-1,m}$,  of the vector wave equation when solved directly in the $ e^{\pm il\theta}$ basis \cite{bhandari}, where  $\theta$ is the azimuthal angle in the cylindrical polar coordinates (this is in sharp contrast to the 
 $\cos l\theta,\sin l\theta$ basis, traditionally used within the optics community, which by and large treats wave propagation in terms of even and odd modes as in  \cite{snyder}; in this basis, the OAM solutions are subsequently constructed through linear combinations of the even $\cos(l\theta)$ and odd $\sin(l\theta)$ solutions; see, e.g., \cite{ram,rusch, huang}).  In what follows, we suppress the radial index $m$ for convenience; however, all numerical calculations assume $m=1$.
The transverse fields of the vector OAM modes, $HE_{l+1}$ and $EH_{l-1}$, denoted  $\vec{e}^{(l+1)}_t$ and  $\vec{e}^{(l-1)}_t$, respectively,
 satisfy the  wave equation 
\begin{equation}
(\nabla^2_t +k^2n^2(r))\vec{e}^{(l\pm1)}_t+\vec{\nabla_t}(\vec{e}^{(l\pm1)}_t.\vec{\nabla_t} ln (n^2))=\beta_{l\pm1}^2 \vec{e}^{(l\pm1)}_t,
\end{equation}
where the third term on the left-hand-side incorporates all the \emph{polarization effects} ; 
 $\beta_{l+1}$ and $\beta_{l-1}$ are, respectively,  the corresponding propagation constants that are determined from two different characteristic equations  , one corresponding to the $HE_{l+1}$ modes and the other to the $EH_{l-1}$ modes \cite{snyder};  $k=2\pi/\lambda$, where $\lambda$ is the wavelength. We assume here  that   refractive index $n$ is given by the familiar expression: $n^2=n_1^2(1-2\Delta f(r))$, where $n_1,~n_2,~f(r),~\Delta(=(n_1^2-n_2^2)/(2n_1^2))$ are respectively the  refractive index of the core, the refractive index of the cladding,   index profile function, and the  profile height parameter (as in the commercial step-index fiber or the currently investigated ring-core fiber \cite {willner, rusch, rusch2}; see Fig. 1). Thus, when   $n_2 \rightarrow n_1, ~\Delta \rightarrow 0$, rendering $\vec{\nabla_t}ln(n^2)=0$ in Eq. 1; in other words,  the polarization effects disappear in the limit $n_2 \rightarrow n_1$, and Eq. 1 reduces to a  scalar wave equation:
\begin{equation}
(\nabla^2_t +k^2n^2(r))\vec{\tilde e}^{(l)}_t=\tilde\beta_{l}^2\vec{\tilde e}^{(l)}_t,
\end{equation}
where $\vec{\tilde e}^{(l)}_t=F_{l}(r)e^{ il\theta}\vec{\epsilon}_{\pm}$ is a derivative of   $\vec{e}^{(l\pm1)}_t$  in the limit $n_2 \rightarrow n_1$ (see Appendix A), and $\tilde\beta_{l} $ is the corresponding scalar propagation constant (in this limit the two different characteristic equations for the vector modes reduce to a single  characteristic equation for the scalar OAM mode with topological charge $l$ and propagation constant $\tilde{\beta}_l$). 
For the general case,  $\Delta <1$, we can make an expansion of $ln(n^2)$ in the parameter $\Delta$. In the \emph{weakly guiding approximation} (WGA), where $\Delta <<1$, expansion to first order in $\Delta$ suffices, which then yields $\vec{\nabla_t} ln (n^2)=-2\Delta (\partial f(r)/\partial r)\hat{r}$ in Eq. 1.
We therefore expect the  polarization correction to the propagation constant 
\begin{equation}
\delta\beta_{l\pm1}^{2}=\beta_{l\pm1}^2 - \tilde\beta_{l}^2
\end{equation}
and the field correction
\begin{equation}
 \delta \vec{e}^{(l\pm1)}_t = \vec{e}^{(l\pm1)}_t -\vec{\tilde e}^{(l)}_t
\end{equation}
to be each of order $\Delta$. We  refer to this field correction as  \emph{polarization-induced} due to its dependence on the parameter $\Delta$.
Analytic expressions up to order $\Delta$  have been given for Eq. 3 in \cite{snyder} (in the context of the scalar  $LP$ modes) and by \cite{volyar, alex, bhandari2} (in the context of scalar OAM modes); specifically in \cite{bhandari2}, explicit expressions for polarization correction to the scalar propagation constant to first order for the step-index fiber are given (note that the polarization corrections for the OAM modes and the corresponding $LP$ modes are identical because they are the solutions of the same scalar wave equation, and the scalar OAM modal fields and the $LP$ modes are linearly related).   Regarding Eq. 4, a correction to order  $\Delta$ in the context of $LP$ modes has also been cited in \cite{snyder}, but no analytic expression is provided therein.
The field correction can also be determined  in perturbation theory \cite{bhandari, bhandari4, ram2}, but the obtained expression is an approximation expressed  in terms of the fields of the traditional scalar modes.
%
\\\\
In this paper, we derive  the exact analytic expressions for the vector fields in multilayered fibers such as the step-index fiber and the ring-core fiber (Fig. 1), and subject them to close scrutiny for their OAM contents.  The structure of the polarization-induced OAM field is especially studied. Subsequently, we ascertain its impact on the polarization correction to the propagation constant of the scalar modes. We begin in Section 2 with the basic definitions  and concepts, constructing in the process the general OAM composition of the vector fields of a  circular fiber, along with the  related concept  of the total angular momentum, $J=L+S$. Subsequently (in Section 3), we derive an analytic form of the vector field solutions  for the step-index fiber, which is then examined for its OAM contents and features;    the extracted features are  then numerically simulated and validated. This analysis is followed  by a discussion of the  extension  to the multilayered fiber. The ramifications of the findings are explored in Section 4 for polarization corrections, leading to a general expression for a multilayered fiber (step-index or the three-layered ring-core fiber). This general expression includes the explicit impact of the polarization-induced field component. Analytic expressions specific to the step-index fiber are subsequently derived for illustrative purposes and numerically simulated.   Section 5 is the summary and discussion.
\section{General OAM composition  of the vector fields in a fiber}
%
%
We begin by  writing the  generic form of the  vector fields expressed directly  in the $e^{\pm il\theta}$ basis used in the solution of the vector wave equation:
\begin{equation}
 \vec{e}^{(l)}=\Big(e^{(l)}_r(r)\hat{r}+e^{(l)}_{\theta}(r)\hat{\theta}+ e^{(l)}_z(r)\hat{z}\Big)e^{il\theta},
\end{equation}
where $l$ is an azimuthal parameter, and where we have assumed that the fiber is translationally invariant and the refractive index depends upon the radial distance $r$ only, i.e., $n^2=n^2(r)$.  Noting that  $\hat{r} = \hat{x}\cos\theta + \hat{y}\sin\theta$ and $\hat{\theta}=-\hat{x}\sin\theta+\hat{y}\cos\theta$, the  transverse field comprised of the first two terms in Eq. 5 can be reexpressed as
\begin{equation} 
\vec{e}^{(l)}_t=(1/{\sqrt{2})}\Big[\vec{\epsilon}_+ e^{i(l-1)\theta}(e^{(l)}_r(r)-ie^{(l)}_\theta(r)) + \vec{\epsilon}_- e^{i(l+1)\theta}(e^{(l)}_r(r)+ie^{(l)}_\theta(r))\Big],
\end{equation}
 where $\vec{\epsilon}_{\pm}=(1/\sqrt{2})(\hat{x}\pm i\hat{y})$  are left/right-circular polarizations associated with a  spin angular momentum $S_z=\pm1$ (left/right circularly polarized photons have spins $\pm\hbar$). 
 Invoking the OAM operator, $L^{op}_z=(1/i)(\partial/\partial\theta)$ \cite{landau}, we observe that in the above field equation, spins of $S_z=+1$ and $S_z=-1$ are coupled with orbital angular momentum values of $L_z=\ell-1$ and $L_z=\ell+1$, respectively so the total angular momentum, denoted $J_z$, in each of the first two terms is $J_z=L_z+S_z = \ell$. Thus, the transverse fields may be regarded as the eigenstates of the total angular momentum operator $J_z$. The second term differs in topological charge by $\pm2$. 
 Changing $l$ to $l\pm1$  then yields 
\begin{equation}
\vec{e}^{(l\pm1)}_t=(1/{\sqrt{2})}\Big[\vec{\epsilon}_{\pm} e^{il\theta}\big(e^{(l\pm 1)}_r(r)\mp i e^{(l\pm 1)}_\theta(r)\big) + \vec{\epsilon}_{\mp} e^{i(l\pm2)\theta}\big(e^{(l\pm 1)}_r(r)\pm i e^{(l\pm 1)}_\theta(r)\big)\Big].
\end{equation}
%
Similar  results are obtained for the associated magnetic field, denoted $\vec{h}_t^{(l\pm1)}$:
\begin{equation} 
\vec{h}^{(l\pm1)}_t=(1/{\sqrt{2})}\Big[\vec{\epsilon}_{\pm} e^{il\theta}\big(h^{(l\pm 1)}_r(r)\mp i h^{(l\pm 1)}_\theta(r)\big) + \vec{\epsilon}_{\mp} e^{i(l\pm2)\theta}\big(h^{(l\pm 1)}_r(r)\pm i h^{(l\pm 1)}_\theta(r)\big)\Big].
\end{equation}
These solutions in conjunction with the $z$ field components, $\vec{e}_z^{(l\pm 1)}$ and $\vec{h}_z^{(l\pm1)}$, are identified with the familiar vector modes, $HE_{l+1}$ and $EH_{l-1}$, expressed in the exponential basis \cite{bhandari3}.
%
%
%
\begin{figure}[htbp]
\vspace{-2mm}
  \centering
  \includegraphics[width=10cm]{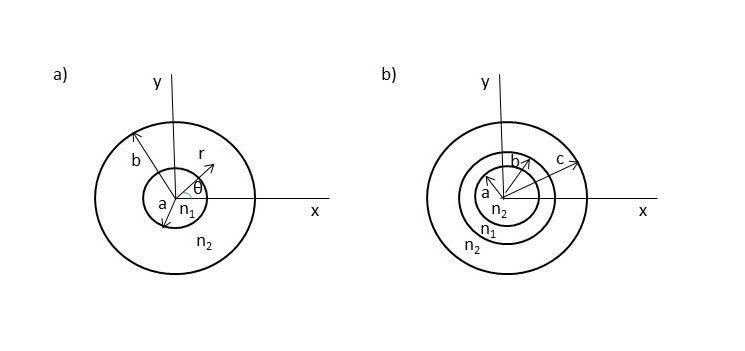}
\vspace{-3mm}
 \caption  {a)  Cross section of a step-index fiber with core radius $a$ and cladding radius $b~ (>>a); ~n_1$ and $n_2$ are the refractive indices of the core and cladding, respectively; $r$ and $ \theta$ along with the $z$ coordinate ($z$ axis coincident with the fiber axis) constitute the polar coordinates; the mode is assumed propagating in the $+z$ direction (out of the plane of the paper).  b)  Cross section of a ring core fiber, with refractive index $n_1$ for the ring core ($a\le r\le b$) and cladding refractive index equal to $n_2<n_1$ in the fiber region: $0\le r\le a$ and $b\le r\le c~(c>>b)$; in the limit $a\rightarrow 0$, we obtain the  step-index fiber.}
\end{figure}
\\\\
%
%
%
In what follows, we determine  the exact forms of the two components in Eq. 7 for the step-index fiber and the ring-core fiber (see Fig. 1) using the  analytic expressions for the radial and the azimuthal components of the electric fields. In particular, we study the OAM characteristics of the fields in a step-index fiber in detail.
\section{OAM modal fields in a multilayered fiber}
%
%
We consider the step-index fiber (two layers) first followed by an extension of  the technique to more than two-layers.
\subsection{Step-Index fiber}
In a standard manner of solving for the fields within a step-index fiber \cite{yariv}, we consider the steady-state solutions: $\vec{E}^{(l)}(\vec{r}, t)=\vec{e}^{(l)}(r,\theta)e^{i(\beta_l z-\omega t)}$ and $\vec{H}^{(l)}(\vec{r}, t)=\vec{h}^{(l)}(r,\theta)e^{i(\beta_l z-\omega t)}$, where $\beta_l$ is the propagation constant  of the mode and $\omega =k/\sqrt{\epsilon_0 \mu_0}$ is the angular frequency. We subsequently apply the Maxwell's equations, starting with
%
%
%
$E_z^{(l)}(r,\theta) = e_z^{(l)}(r)e^{il\theta}$, where $e_z^{(l)}(r)=A_lJ_l(p_lr)$, and $H_z^{(l)}(r,\theta) = h_z^{(l)}(r)e^{il\theta}$, where $h_z^{(l)}(r)=B_lJ_l(p_lr)$, within the fiber  core of radius $a$; similarly, for the cladding region ($r\ge a)$, we write $E_z^{(l)}(r,\theta) = e_z^{(l)}(r)e^{il\theta}$, where $e_z^{(l)}(r)=C_lK_l(q_lr)$, and $H_z^{(l)}(r,\theta) = h_z^{(l)}(r)e^{il\theta}$, where $h_z^{(l)}(r)=D_lK_l(q_lr)$.  The results for the radial profiles, $e_r^{(l)}$ and $e_\theta^{(l)}$, comprising the transverse field (see Eq. 5) are
\begin{equation}
e^{(l)}_r=(i\beta_l/ p_l^2)\Big[A_lp_lJ_l'(p_lr)+B_l(il/r)(\mu_0\omega/\beta_l )J_l(p_lr)\Big],
\end{equation}
\begin{equation}
e^{(l)}_\theta=(i\beta_l /p_l^2)\Big[A_l(il/r)J_l(p_lr)-(\mu_0\omega p_l/\beta_l )B_lJ_l'(p_lr)\Big];~~~~r\le a
\end{equation}
%
\begin{equation}
e^{(l)}_r=-(i\beta_l/ q_l^2)\Big[C_lq_lK_l'(q_lr)+D_l(il/r)(\mu_0\omega/\beta_l )K_l(q_lr)\Big],
\end{equation}
\begin{equation}
e^{(l)}_\theta=-(i\beta_l /q_l^2)\Big[C_l(il/r)K_l(q_lr)-(\mu_0\omega q_l/\beta_l )D_lK_l'(q_lr)\Big].~~~~r\ge a
\end{equation}
$J_l$ and $K_l$ are the Bessel and the modified Bessel functions, respectively. $A_l,B_l,C_l,D_l$ are constants that are determined from  matching  the tangential  components, $e^{(l)}_{\theta}, e^{(l)}_z, h^{(l)}_{\theta}, h^{(l)}_z$, at the interface $r=a$; $ kn_2\le\beta_l\le kn_1;$ $p^2_l= k^2n_1^2 -\beta_l^2 (\ge 0); q_l^2=-(k^2n_2^2-\beta_l^2) (\ge 0)$;     $k^2=\omega^2\mu_0\epsilon_0$.    Physically, $p_l$ is the transverse component of the  propagation constant (wave number), $kn_1$, within the core medium, while $\beta_l$ is the horizontal component, coincident withe the z axis; parameter $q_l$ has a similar interpretation, being the magnitude of the imaginary transverse component within the cladding, where the fields quickly decline to zero.    
 For the fields within the core, we now find using the identities, $J_{l\pm1}(x)=\mp J'_l(x)+(l/x)J_l(x)$ , that 
\begin{equation}
e^{(l)}_r\mp ie^{(l)}_{\theta}=\pm(i\beta_l/p_l)A_lJ_{l\mp1}(p_l r)\Big(1\pm(i\omega \mu_0/\beta_l)(B_l/A_l)\Big),
\end{equation}
and similarly for the fields within the cladding ($r\ge a$) we obtain
\begin{equation}
e^{(l)}_r\mp ie^{(l)}_{\theta}=(i\beta_l/q_l)C_lK_{l\mp1}(q_l r)\Big(1\pm(i\omega \mu_0/\beta_l)(D_l/C_l)\Big);
\end{equation}
$D_l/C_l=B_l/A_l$ and $C_l/A_l=J_l(p_la)/K_l(q_la)$\cite{yariv}. 
\subsubsection{Analysis of the $J=l+1$ case}
Changing $l$ to $l+1$ in Eq. 13 and inserting the resulting expressions in Eq. 7, we obtain for the fields within the core
\begin{equation}
\vec{e}_t^{(l+1)}=  (1/\sqrt{2})   (i\beta_{l+1}/p_{l+1})A_{l+1}\Big[\gamma^{(l+1)}_+ J_l(p_{l+1} r)e^{il\theta}\vec{\epsilon_+}+\gamma^{(l+1)}_- J_{l+2}(p_{l+1} r)e^{i(l+2)\theta}\vec{\epsilon_-}\Big],
\vspace{-2mm}
\end{equation}
where 
\begin{equation}
\gamma^{(l+1)}_\pm = \pm\Big(1\pm (i\omega\mu_0/\beta_{l+1})(B_{l+1}/A_{l+1})\Big),
\end{equation}
and $B_{l+1}/A_{l+1}$ is determined from the boundary conditions as discussed earlier. 
\\\\
Eq. 15 is a very general result, and a remarkable one because the spatial forms in Eq. 15 resemble the scalar field amplitudes (solutions of the scalar wave equation). Parameter $p_{l+1}$ in the argument of the Bessel functions, $J_l$ and $J_{l+2}$, is given by $p^2_{l+1}=k^2n_1^2-\beta_{l+1}^2$. 
The corresponding counterpart $\tilde{p}_{ l}$ in the solution of the scalar wave equation, $\vec{\tilde e}^{(l)}_t\sim J_{l}(\tilde{p}_l r)e^{ il\theta}\vec{\epsilon}_{+}$ (see Eq. 2), is similarly related by $\tilde{p}_l^2=k^2n_1^2-\tilde{\beta}_l^2$. Consequently, $\tilde{p}_l^2- p_{l+1}^2=\beta_{l\pm1}^2 - \tilde\beta_{l}^2= \delta\beta_{l\pm1}^2$, which is of order $\Delta <<1$ in WGA and therefore very small.  Thus,  $p_{l+1}$ in the argument of $J_l$ in Eq. 15 can be replaced with $\tilde{p}_{ l}$, implying \emph{we may identify the first term in Eq. 15 with the field of the  traditional scalar $OAM_l$ mode possessing    a topological charge $l$ and     an amplitude $O_{l}(r,\theta) =J_{l}(\tilde{p}_l r)e^{ il\theta} $ \cite{bhandari}} (see also Appendix A). Peak intensity defined as the square of the magnitude of $O_{l}$ occurs at the first maximum of the $J_l (\tilde{p}_l r)$ function. If we define $r=r^{(l)}_{max}$ as this maximum, then $r^{(l)}_{max}=x_{max}/\tilde{p}_l$, where $x_{max}$ is the first maximum of $J_l(x), l>0$. Since the Bessel function rises from its zero value with  increase in its argument value, the intensity of this mode as a function of the radial distance $r$ is the familiar donut pattern, often observed in experimental data (see, e.g, \cite{huang}). 
\\\\
\emph{The second term  in Eq. 15, however, does \emph{not} correspond to the field of the  traditional scalar $OAM_{l+2}$ mode because its radial profile, $J_{l+2}(p_{l+1}r)\approx J_{l+2}(\tilde{p}_{l}r)$ is not the same as  the profile, $J_{l+2}(\tilde{p}_{l+2}r)$ of the traditional $OAM_{l+2}$ scalar mode}. This is due to the fact that $\tilde{p}_{l+2}\ne \tilde{p}_{l}$. In fact, $\tilde{p}_{l+2}> \tilde{p}_{l} $ (see Table 1) because $\tilde{\beta}_{l+2}< \tilde{\beta}_{l}$. As a result, the second term in Eq. 15 is a 
\begin{table}
\centering
\caption{The second term in Eq. 15 is the field of a nontraditional (modified)  scalar $OAM_{l+2}$ mode, whose intensity maximum at the radial distance, $r'^{(l+2)}_{max}$ is shifted outward  by a factor $f^+_l= \tilde{p}_{l+2}/\tilde{p}_{l}$ relative to the position $r^{(l+2)}_{max}$ of the intensity maximum of the conventional scalar $OAM_{l+2}$ mode; the numerical values are for a step-index fiber with parameters: $n_1=1.461,~n_2=1.444,~ a=25 \mu m,\lambda=1.55 \mu m$ (see Fig. 1).} 
\begin{tabular}{ |c|c|c|c|c|c|} 
 \hline
$ l$&$\tilde{p}_l$&$\tilde{p}_{l+2}$&$f^+_l$&$r^{(l+2)}_{max}/a$&$ r'^{(l+2)}_{max}/a$\\ 
\hline
 1&3.6681&6.1063& 1.66&0.69&1.15\\ 
\hline
 2&4.9158&7.2617&1.48&0.73&1.08\\ 
 \hline
3&6.1063&8.3927&1.37&0.76&1.04\\
\hline
4&7.2617&9.5056&1.31&0.79&1.03\\
\hline
5&8.3927&10.6043&1.26&0.81&1.02\\
\hline
\end{tabular}
\end{table}
%
\begin{footnotesize}
\begin{figure}[htbp]
  \centering
  \begin{subfigure}[b]{0.35\linewidth}
    \includegraphics[width=\linewidth]{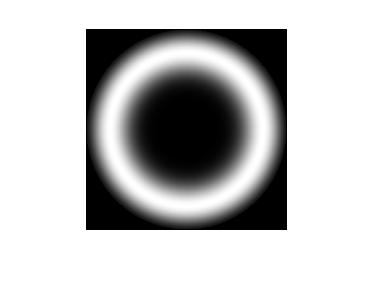}
  \end{subfigure}
 \begin{subfigure}[b]{0.35\linewidth}
    \includegraphics[width=\linewidth]{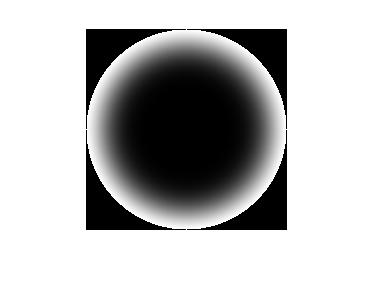}
  \end{subfigure}
 \begin{subfigure}[b]{0.35\linewidth}
    \includegraphics[width=\linewidth]{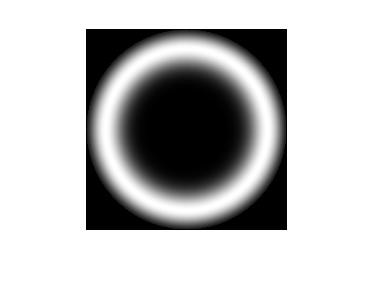}
  \end{subfigure}
\begin{subfigure}[b]{0.35\linewidth}
    \includegraphics[width=\linewidth]{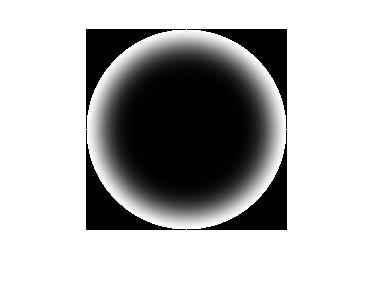}
  \end{subfigure}
  \label{multimode}
\caption  {The   $J=l+1$ case;  the intensity pattern within the core, $J^2_{l+2}(\tilde{p}_{l+2}r)$,   corresponding to the traditional scalar $OAM_{l+2}$ mode (on the left) and the  expanded out intensity pattern ($J_{l+2}^2(\tilde{p}_lr)$) corresponding to the modified scalar $OAM_{l+2}$ modal field (on the right), as predicted by our theoretical expression (the second term in Eq. 15); see text for more details; the top row is for $l=3$ and the bottom row is for $l=5$ (see Table 1).}
\end{figure}
\end{footnotesize}
 \emph{modified} $OAM_{l+2}$ modal field, whose predicted intensity maximum occurs  at $r=r'^{(l+2)}_{max}$, where
\begin{equation}
r'^{(l+2)}_{max}=(\tilde{p}_{l+2}/\tilde{p}_{l})r^{(l+2)}_{max};
\end{equation}
 $r^{(l+2)}_{max}$ is the radial distance at which the intensity of the traditional $OAM_{l+2}$ mode reaches its maximum.  Because $\tilde{p}_{l+2}> \tilde{p}_{l} $, this maximum is shifted outward in the characteristic donut shaped intensity pattern, even beyond $r=a$ (see Table 1). However,  because the fields drop off exponentially into the cladding region (being given by Eq. 14 for $r>a$), the  maximum consequently occurs at $r=a$. The relatively low intensity region (the dark region of the donut), nevertheless, scales by a factor of $f^+_l=\tilde{p}_{l+2}/\tilde{p}_{l}>1$.
%
Figure 2 shows the predicted intensity pattern (within the core) of the modified scalar OAM modal field (on the right)  contrasted with the corresponding  traditional scalar mode (on the left) for $l=3$ (top row) and $l=5$ (bottom row); the the step-index fiber is described by  $n_1=1.461,~ n_2=1.444$, core radius $a=25~\mu m$; wavelength $\lambda=1.55 \mu m$ is assumed.  
\begin{footnotesize}
\begin{figure}[htbp]
  \centering
  \begin{subfigure}[b]{0.35\linewidth}
    \includegraphics[width=\linewidth]{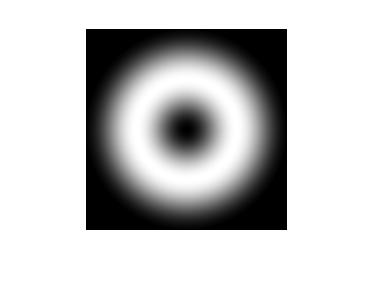}
  \end{subfigure}
 \begin{subfigure}[b]{0.35\linewidth}
    \includegraphics[width=\linewidth]{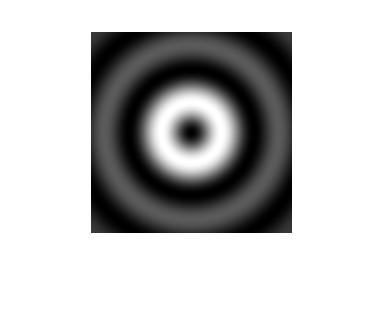}
  \end{subfigure}
 \begin{subfigure}[b]{0.35\linewidth}
    \includegraphics[width=\linewidth]{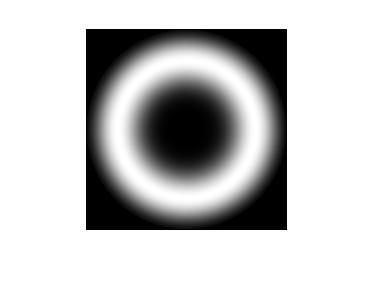}
  \end{subfigure}
\begin{subfigure}[b]{0.35\linewidth}
    \includegraphics[width=\linewidth]{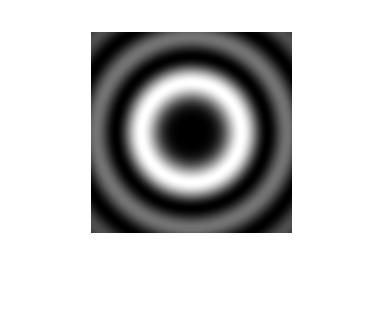}
  \end{subfigure}
  \label{multimode}
\caption  {The   $J=l-1$ case;  the intensity pattern within the core,  $J_{l-2}^2(\tilde{p}_lr)$,   corresponding to the traditional scalar $OAM_{l-2}$ mode (on the left) and the  shrunken intensity pattern $J^2_{l-2}(\tilde{p}_{l-2}r)$ corresponding to the modified scalar $OAM_{l-2}$ field (on the right), as predicted by our theoretical expression (the second term in Eq. 18); see text for more details; the top row is for $l=3$ and the bottom row is for $l=5$ (see Table 2).}
\end{figure}
\end{footnotesize}
%
%
%
%
%
\subsubsection{Analysis of the $J=l-1$ case}
Changing $l$ to $l-1$ in Eq. 15 and substituting in Eq. 7, we obtain
\begin{equation}
\vec{e}_t^{(l-1)}=  (1/\sqrt{2})   (i\beta_{l-1}/p_{l-1})A_{l-1}\Big[\gamma^{(l-1)}_+ J_l(p_{l-1} r)e^{il\theta}\vec{\epsilon_-}+\gamma^{(l-1)}_- J_{l-2}(p_{l-1} r)e^{i(l-2)\theta}\vec{\epsilon_+}\Big],
\end{equation}
where
\begin{equation}
\gamma^{(l-1)}_\pm =   \mp   \Big(1  \mp    (i\omega\mu_0/\beta_{l-1})(B_{l-1}/A_{l-1})\Big).
\end{equation}
The field corresponds to the $EH_{l-1}$ mode. In WGA, $\beta_{l-1}\approx \tilde{\beta}_l$, the corresponding solution of the scalar wave equation (see Section 1); consequently, $p_{l-1}\approx \tilde{p}_l$. So, the first term corresponds to the traditional scalar mode with the radial profile, $J_{l}(\tilde{p}_{l}r)$, while the second term is a \emph{modified} scalar modal field because the associated radial function, $J_{l-2}(p_{l-1}r)\approx J_{l-2}(\tilde{p}_l r)$,  is different from  $J_{l-2}(\tilde{p}_{l-2})r$, the radial profile of the scalar OAM mode characterized by topological charge $l-2$ . The donut  intensity pattern here has a  radius 
\begin{equation}
r'^{(l-2)}_{max}=(\tilde{p}_{l-2}/\tilde{p}_l)r^{(l-2)}_{max},
\end{equation}
which is reduced since $\tilde{p}_{l-2}<\tilde{p}_l$, i.e., an intensity pattern as determined solely by the  modified modal field, would be shifted  inward  towards the center by a factor $f^-_l\approx \tilde{p}_{l-2}/\tilde{p}_{l}$; see Table 2 and Figure 3; \emph{the intensity pattern here displays the secondary maxima not present in the traditional scalar $OAM_{l-2}$ fields}.
\begin{table}
\centering
\caption{The second component in Eq. 15 is a nontraditional (modified)  scalar $OAM_{l-2}$ modal field, whose intensity maximum, $r'^{(l-2)}_{max}$ is shifted by a factor $f^-_l= \tilde{p}_{l-2}/\tilde{p}_{l};~$$r^{(l-2)}_{max}$ is the position of the intensity maximum of the conventional scalar $OAM_{l-2}$ mode; the numerical values are for a step-index fiber with parameters: $n_1=1.461,~n_2=1.444,~ a=25 \mu m,\lambda=1.55 \mu m$ (see Fig. 1).} 
\begin{tabular}{ |c|c|c|c|c|c|} 
 \hline
$ l$&$\tilde{p}_l$&$\tilde{p}_{l-2}$&$f^-_l$&$r^{(l-2)}_{max}/a$&$ r'^{(l-2)}_{max}/a$\\ 
\hline
 3&6.1063&3.6681&0.60&0.50&0.30\\ 
 \hline
4&7.2617&4.9158&0.68&0.62&0.42\\
\hline
5&8.3927&6.1063&0.73&0.69&0.50\\
\hline
6&9.5056&7.2617&0.76&0.73&0.56\\
\hline
7&10.6043&8.3927&0.79&0.76&0.60\\
\hline
\end{tabular}
\end{table}
\subsubsection{Relative amplitude of the modified scalar modal field}
We need to evaluate the coefficients, $\gamma^{(l+1)}_\pm$ (see Eqs. 15 and 16) and $\gamma^{(l-1)}_\pm$ (see Eqs. 18 and 19). We first note that 
\begin{equation}
\frac{B_{l}}{A_{l}}=\frac{il\beta_{l}}{\omega\mu_0}\Big(\frac{1}{q^2_{l}a^2}+\frac{1}{p_{l}^2a^2}\Big)\Big(\frac{J'_{l}(p_{l}a)}{p_{l}aJ_{l}(p_{l}a)}+\frac{K'_{l}(q_{l}a)}{q_{l}aK_{l}(q_{l}a)}\Big)^{-1},
\end{equation}
which is easily derived from the boundary conditions at the core-cladding interface \cite{yariv}. The characteristic equation for vector modes is well-known \cite{snyder, yariv}:
\begin{equation}
\Big(n_1^2\frac{J'_l(p_la)}{p_la J_l(p_la)}+n_2^2\frac{K'_l(q_la)}{q_laK_l(q_la)}\Big)\Big(\frac{J'_l(p_la)}{p_lJ_l(p_la)}+\frac{K'_l(q_la)}{q_laK_l(q_la)}\Big)=l^2\Big(\frac{1}{q_l^2a^2}+\frac{1}{p_l^2a^2}\Big)\frac{\beta_l^2}{k^2}.
\end{equation}
 In the  limit, $n_2\rightarrow n_1$, 
the above characteristic equation reduces to the  characteristic equation for the scalar modes \cite{snyder,yariv}, assuming $\beta_l\approx kn_1$. In a real fiber, $n_2$ and $n_1$ are not exactly equal, being related by  $n_2^2=n_1^2(1-2\Delta)$, where $\Delta$ is usually a small fraction. Inserting this expression in Eq. 22 then leads to the following relationship:
\begin{equation}
\delta_l=\pm\frac{k n_1\alpha_l}{\beta_l l}\Big(1-\frac{2\Delta\kappa_l}{\alpha_l}\Big)^{1/2},
\end{equation}
where
\begin{equation}
\delta_l=\Big(\frac{1}{q_l^2a^2}+\frac{1}{p_l^2a^2}\Big),
\end{equation}
\begin{equation}
\alpha_l=\Big(\frac{J'_l(p_la)}{p_laJ_l(p_la)}+\frac{K'_l(q_la)}{q_laK_l(q_la)}\Big),
\end{equation}
\begin{equation}
\kappa_l=\frac{K'_l(q_la)}{q_laK_l(q_la)}.
\end{equation}
In terms of the above parameters, Eq. 21 can  be rewritten as
\begin{equation}
\frac{B_l}{A_l}=\frac{il\beta_{l}}{\omega\mu_0}\Big(\frac{\delta_l}{\alpha_l}\Big),
\end{equation}
which further upon substitution of Eq. 23 yields
\begin{equation}
\frac{B_l}{A_l}=\pm \frac{ikn_1}{\omega\mu_0}\Big(1-\frac{2\Delta\kappa_l}{\alpha_l}\Big)^{1/2}.
\end{equation}
\emph{This is a very general and exact result, regardless of the magnitude of $\Delta$}. Eq. 28, when substituted in Eq. 13, determines the complete analytic expressions for the electric field. The + sign corresponds to the $EH_{l-1}$ mode and the negative sign to the $HE_{l+1}$ mode. In the limit $\Delta \rightarrow 0,~B_l/A_l=\pm ik n_1/(\omega \mu_0)$, which leads to the conventional scalar modal fields (see Appendix A).
\\\\
\underline{$J=l+1$}
\\\\
We change $l$ to $l+1$ in Eq. 28 and insert  the resulting expression for $B_{l+1}/A_{l+1}$  with a negative sign,  in Eq. 16 to obtain
\begin{equation}
\gamma^{(l+1)}_+= 1+ \frac{kn_1}{\beta_{l+1}}\Big(1-\frac{2\Delta\kappa_{l+1}}{\alpha_{l+1}}\Big)^{1/2},
\end{equation}
\begin{equation}
\gamma^{(l+1)}_-= -1+ \frac{kn_1}{\beta_{l+1}}\Big(1-\frac{2\Delta\kappa_{l+1}}{\alpha_{l+1}}\Big)^{1/2}.
\end{equation}
When  $\Delta ~<<1$  (e.g., $\Delta \approx 0.01$ in a standard commercial multimode step-index fiber  and $ \approx 0.02$ in considered  ring-core fibers\cite{rusch2,willner}), we can expand Eqs. 29 and 30 to first order in $\Delta$; further, using the fact that $\beta_{l+1}\approx kn_1$, we arrive at 

\begin{equation}
\gamma^{(l+1}_+= 2-\Delta\Big(\frac{\kappa_{l+1}}{\alpha_{l+1}}\Big)
\end{equation}
and
\begin{equation}
\gamma^{(l+1)}_-=-\Delta\Big(\frac{\kappa_{l+1}}{\alpha_{l+1}}\Big),
\end{equation}
leading to the relative amplitude
\begin{equation}
\mu_{l+1}=\gamma^{(l+1)}_-\big/\gamma^{(l+1)}_+=-\Delta\Big(\frac{\kappa_{l+1}}{2\alpha_{l+1}}\Big)
\end{equation}
   to first order in $\Delta$. Eq. 33 provides the relative strength of the modified scalar modal field, which we find to be of order $\Delta$ in the first order correction; this is consistent with the expansion of the electric field  in terms of $\Delta$ in \cite{snyder}. This modified scalar modal field is  the \emph{polarization-induced} component of the vector field \cite{bhandari2}, as discussed in Section 1. Insertion of Eqs. 31 and 33 in Eq. 15 then provides a complete description of  the transverse vector OAM field in terms of its two components differing in topological charge by $2$. The parameters $p_{l+1}$ and $q_{l+1}$ in the expressions for $\kappa_{l+1}$ and $\alpha_{l+1}$ correspond to the vector mode solution, $HE_{l+1}$. The two components in Eq. 15 move as one entity (in conjunction with the z component, which is very small; see Appendix A) with a propagation constant   $\beta_{l+1}$ corresponding to the $HE_{l+1}$ vector mode.    In other words, the polarization-induced field, the second component, which has the same propagation constant $\beta_{l+1}$ as the first (primary) component, must have the same value $p_{l+1} (\approx \tilde{p}_l$ in WGA) as the primary component, and not $\tilde{p}_{l+2}$, which is characteristic of the traditional scalar mode of topological charge $l+2$ in WGA.    
\\\\
\underline{$J=l-1$}
\\\\
   Here we change $l$ to $l-1$ in Eq. 28 and insert the resulting expression (with the positive sign) in Eq. 19. This yields the following expressions: 
\begin{equation}
\gamma^{(l-1)}_+= -1- \frac{kn_1}{\beta_{l+1}}\Big(1-\frac{2\Delta\kappa_{l+1}}{\alpha_{l+1}}\Big)^{1/2},
\end{equation}
\begin{equation}
\gamma^{(l-1)}_-= 1- \frac{kn_1}{\beta_{l+1}}\Big(1-\frac{2\Delta\kappa_{l+1}}{\alpha_{l+1}}\Big)^{1/2}.
\end{equation}
Retaining the first term in the binomial expansion of Eqs. 34 and 35 in $\Delta (<<1)$, and  using the fact that $\beta_{l-1}\approx kn_1$, we arrive at 

\begin{equation}
\gamma^{(l-1}_+= -2+\Delta\Big(\frac{\kappa_{l+1}}{\alpha_{l+1}}\Big)
\end{equation}
and
\begin{equation}
\gamma^{(l-1)}_-=\Delta\Big(\frac{\kappa_{l+1}}{\alpha_{l+1}}\Big),
\end{equation}
which implies that
the  relative  polarization-induced field coefficient, denoted $\mu_{l-1}$, here is given  by
\begin{equation}
\mu_{l-1}=\gamma^{(l-1)}_-\big/\gamma^{(l-1)}_+=-\Delta\Big(\frac{\kappa_{l-1}}{2\alpha_{l-1}}\Big);
\end{equation}
parameters $p_{l-1}$ and $q_{l-1}$, occurring in $\kappa_{l-1}$ and $\alpha_{l-1}$, correspond to the solution $EH_{l-1}$, which has a different characteristic equation than that of the corresponding $HE_{l-1}$ mode \cite{snyder}. Insertion of Eqs. 36 and 38 in Eq. 18 now yields a complete description of the contribution of the polarization-induced component.    This polarization-induced component of topological charge $l-2$ moves with the same propagation constant $\beta_{l-1}$ as the primary component  (which is the field  of a traditional scalar mode in WGA), and therefore is characterized by $p_{l-1}\approx \tilde{p}_l$, and not $\tilde{p}_{l-2}$, the parameter corresponding to the traditional scalar mode of topological charge $l-2$.    
  \subsection{Numerical validation}
Here we compare the exact solution of the vector wave equation with the results derived on the basis of $\Delta<<1$; in the latter case, we recast the results in terms of the solutions of  the scalar wave equation. Consequently, focusing first on the $J=l+1$ case, we 
insert Eqs.  31 and 32 (or 33) into Eq. 15, and set $\beta_{l+1}\approx \tilde{\beta}_l$, $p_{l+1} \approx \tilde{p}_l$, and $q_{l+1}\approx \tilde{q}_l $, results valid in WGA.   Subsequently, we arrive at
\begin{equation}
|\vec{e}_t^{(l+1)}|^2=2(\tilde{\beta}_l/\tilde{p}_l )^2|A_{l+1}|^2(1-\Delta\tilde{\kappa}_{l+1}/(2\tilde{\alpha}_{l+1}))^2\Big[J^2_l(\tilde{p}_lr)+\frac{\Delta^2\tilde{\kappa}^2_l}{4\tilde{\alpha}^2_l}J^2_{l+2}(\tilde{p}_l r)\Big],
\end{equation}
which, apart from the constant, $A_{l+1}^2$,  is an expression given entirely in terms of the scalar parameters (the second term is the contribution of the polarization-induced component).  This expression is to be contrasted with the expression 
\begin{equation}
|\vec{e}_t^{(l+1)}|^2=|e_r^{(l+1})|^2+|e_{\theta}^{(l+1)}|^2,
\end{equation}
where $e_r^{(l+1)}$ and $e_{\theta}^{(l+1)}$ represent the exact vector wave equation solutions given by Eqs. 9 and 10, but with $l$ replaced with $l+1$. We now define $I^{(l+1)}_S = \int_0^a\int_0^{2\pi}|\vec{e}_t^{(l+1)}|^2rdrd\theta=2\pi\int_0^a|\vec{e}_t^{(l+1)}|^2rdr$, which is representative of the intensity of the field;  $a$ is the  radius of the fiber core (where most light energy resides) and $|\vec{e}_t^{(l+1)}|^2$ is given by Eq. 39. Similarly, we write $I^{(l+1)}_V= \int_0^a\int_0^{2\pi}|\vec{e}_t^{(l+1)}|^2rdrd\theta =2\pi\int_0^a|\vec{e}_t^{(l+1)}|^2rdr $, where $|\vec{e}_t^{(l+1)}|^2$ is given by Eq. 40. For the $J=l-1$ case, we use Eqs. 36 and 37 (or 38), in Eq. 18, which yields the same expression as in Eq. 39, except that  the subscript $l+1$ is replaced with $l-1$ and subscript $l+2$ is replaced with $l-2$. We similarly define $I_S^{(l-1)}$ and $I_V^{(l-1)}$. 
\begin{table}
\centering
\caption{$I^{'(J)}_V=I_V/(2\pi|A_{J}|^2)$ and $I^{'(J)}_S=I_S/(2\pi|A_{J}|^2)$ for the $J=l\pm1$ cases and  various values of topological charge $l$; the percentage difference (fourth column), defined as $(I^{'(J)}_V-I^{'(J)}_S)/I^{'(J)}_V x 100$,   is of the order of 0.1 or less in magnitude (see text for more details); $n_1=1.461, n_2=1.444, a =25 \mu m, \lambda=1.55 \mu m$. } 
\begin{tabular}{|c|c|c|c|} 
 \hline
$ l$&$I^{'(l+1)}_V$&$I^{'(l+1)}_S$&\%$diff $\\ 
\hline
2&113.300575&113.476040&-0.15\\ 
 \hline
 3&56.538155&56.611168&-0.13\\ 
 \hline
4&32.310729 &32.342199&-0.10\\
\hline
5&20.196714&20.208784&-0.06\\
\hline
\end{tabular}
\quad
\begin{tabular}{ |c|c|c|c|} 
 \hline
$ l$&$I^{'(l-1)}_V$&$I^{'(l-1)}_S$&\%$diff $\\ 
\hline
2&115.246009&115.372053&-0.11\\ 
 \hline
 3&57.383678&57.424266&-0.07\\ 
 \hline
4&32.797496 &32.805129&-0.02\\ 
\hline
5&20.5140336&20.507828&0.03\\
\hline
 \end{tabular}
\end{table}
\\\\
For our calculations done in MATLAB in double precision, we consider a multimode fiber with the same parameters as those used in Tables 1 and 2. Table 3 shows the results of numerical integration done for $I^{'(J)}_V $ and  $I^{'(J)}_S$, where $J=l\pm 1$; for the former, we used two different characteristic equations, one corresponding to the $HE_{l+1}$ modes and the other to the $EH_{l-1}$ modes, while for $I^{'(J)}_S$, we used the single characteristic equation corresponding to the scalar modes  \cite{snyder}. The differences (fourth table column) are of the order of a tenth percent or less, corroborating the veracity of our results expressed in terms of the scalar parameters.  Furthermore, we observed that, using  parameter $\tilde{p}_{l+2}$, instead of $\tilde{p}_l$, in the argument of $J_{l+2}$ in Eq. 39, causes  \%diff in Table 3 (for the $l+1$ case) to increase by a factor of 2 to 9 in magnitude,  supporting the fact that the polarization-induced component of the electric field is indeed the field of  a modified scalar OAM mode of topological charge $l+2$ as described in Section 3.1, and not the field of a traditional scalar mode of the same topological charge. Similar worsening of the results (although less dramatic) occurs in the $J= l-1$ case, when $\tilde{p}_l$ in the argument of the Bessel function, $J_{l-2}$, is replaced with $\tilde{p}_{l-2}$. 
%
%
%
%
\subsection{Extension to a multilayered fiber}
In extending the above theoretical concepts  to a cylindrical waveguide with more than two layers, 
we observe that, within each layer of the waveguide, the field solutions are, in general, a linear combination of the Bessel functions: $J_l(\alpha_i r)$ and $Y_l(\alpha_i r)$, when $\alpha_i^2=(k^2n_i^2-\beta_l^2)$ is positive,  and a linear combination of the modified Bessel functions: $I_l(\alpha_i r)$ and $K_l(\alpha_i r)$, when $\alpha_i^2$  is negative. 
%
%
For the 3-layer fiber, defined by a ring core of radii $a$ and $b (>a)$, refractive index $n_1$, and inner and outer cladding refractive index of   $n_2<n_1$ (Fig. 1), the z component of the field is expressible in the azimuthal exponential basis as
\begin{equation}
\begin{split}
e^{(l)}_z=&A^{(l)}_1 I_l(q_lr)e^{il\theta}~~~( r\le a),\\
&=[C^{(l)}_1J_l(p_lr)+C^{(l)}_2Y_l(p_lr)]e^{il\theta} ~~~ (a \le r\le b), \\
&=A^{(l)}_2 K_l(q_l r )e^{il\theta}~~~(r\ge b);
\end{split}
\end{equation}
note that the fields within the cladding region die out. Similarly,
\begin{equation}
\begin{split}
h^{(l)}_z&=B^{(l)}_1 I_l(q_lr)e^{il\theta}~~~ (r\le a),\\
&=[D^{(l)}_1J_l(p_lr)+D^{(l)}_2Y_l(p_lr)]e^{il\theta}~~~ (a \le r\le b), \\
&=B^{(l)}_2 K_l(q_l r )e^{il\theta}~~~(r\ge b).
\end{split}
\end{equation}
%
%
Within the ring, $a\le r\le b$, where the fields are primarily concentrated, we obtain, using Maxwell's equations (as in the step-index fiber case) and  the appropriate Bessel function  recursion relations, the following expression:
\begin{equation}
\begin{split}
e^{(l)}_r\mp i e^{(l)}_{\theta}=(i\beta_l /p_l) C^{(l)}_1&[(\pm J_{l\mp1}(p_lr)(1\pm (i\mu_0\omega/\beta) (D^{(l)}_1/C_1))\\
&\pm C^{(l)}_2/C^{(l)}_1(1\pm(i\mu_0\omega/\beta)(D^{(l)}_2/C^{(l)}_2)Y_{l\mp1}(p_lr))];
\end{split}
\end{equation}
 the ratios, $D^{(l)}_1/C^{(l)}_1, C^{(l)}_2/C^{(l)}_1, D^{(l)}_2/C^{(l)}_2$ are  determined
from the boundary conditions at the two interfaces,  $r=a$ and $r=b$. Changing  $l$ to $l+1$
 and inserting the  resulting expression in Eq. 7, for the $l+1$ case the  first term is the linear combination of $J_{l}(p_{l+1}r)$ and $Y_{l}(p_{l+1}r)$ coupled with polarization $\vec{\epsilon}_+$   and the second term is the  linear combination of $J_{l+2}(p_{l+1}r)$ and $Y_{l+2}(p_{l+1}r)$ coupled with polarization $\vec{\epsilon}_-$. This is very similar to the expression, Eq. 15, for the step-index fiber. A detailed analysis, as in the step-index fiber, is cumbersome (as is also seen in \cite{rusch2}) and not done here. However, we note that the fields are confined within the ring and, intuitively, similar conclusions  would hold  as in the step-index fiber, to which the ring-core fiber reduces in the limit $a\rightarrow 0$;  for thin rings, however, the  changes in the radius of the annular bright intensity ring would accordingly be small.  
%

%
%
\section{General  form of the  polarization correction}
In this section, we evaluate the impact of the second component (the modified scalar modal field) on the polarization correction to  the propagation constant of a scalar mode. Prior work \cite{snyder, volyar, alex, bhandari2} has not considered it, partly due to the lack of an analytic expression as discussed in Section 1.
\\\\
%
We first derive a generic form for polarization correction of propagation constant in a mutlilayered fiber.
This form includes the explicit contribution of the polarization-induced component (the second component) in Eq. 7.   We also assume the WGA so that we can write the polarization-induced field correction  (see Eq. 4 and Eq. 7) as 
\begin{equation} 
\delta\vec{e}_t^{(l\pm1)}=G_{l\pm2}(r)e^{i(l\pm2)\theta)}\vec{\epsilon}_\mp.
\end{equation}
 For multilayered fibers such as the step-index fiber and the three-layer ring-core fiber, characterized by a core refractive index of $n_1$ and a cladding refractive index of  $n_2$ (see Fig. 1),  the function $G_{l\pm2}(r)=G_{l\pm2}(p_{l+1}r)$, where the parameter $p_{l+1}^2=k^2n_1^2-\beta_{l+1}^2$. Recalling $ \vec{\tilde{e}}_t^{(l)}=F_{l}(r)e^{il\theta}\vec{\epsilon}_\pm$, we substitute $ \vec{e}^{(l\pm1)}_t= \vec{\tilde e}^{(l)}_t + \delta \vec{e}^{(l\pm1)}_t$ into Eq. 1.
After taking scalar products and performing  the necessary integrations  (see Appendix B), we arrive at the general form for \emph{polarization correction}
\begin{equation}
\begin{split}
\delta\beta^2_{l\pm1}&=\beta_{l\pm1}^2-\tilde{\beta}_{l}^2\\
&=-\pi\Big[\int_0^\infty\frac{\partial (ln(n^2)}{\partial r}F_{l}\big(\frac{dF_{l}(r)}{dr}\mp \frac{l}{r}F_{l}\big)rdr+\int_0^\infty\frac{\partial ln(n^2)}{\partial r}G_{l\pm2}\big(\frac{dF_{l}(r)}{dr}\mp \frac{l}{r}F_{l}\big)rdr\Big].
\end{split}
\end{equation}
The radial amplitudes, $F_l(r)$ and $G_{l\pm2}(r)$, are characteristic of the multilayered fiber under consideration. The first integral is identified with the familiar polarization correction \cite{bhandari2, snyder,volyar}; \emph{the second integral is the correction due to the polarization-induced field component}.  Furthermore, $\Delta$ being less than one  allows us to make an  expansion of $ln(n^2)$ in the parameter $\Delta$ (Section 1). The function $G_{l\pm2}$ in the second term of Eq. 45 is of order $\Delta$, so an expansion of $ln (n^2)$ to first order in $\Delta$ yields a result quadratic in $\Delta$. 
%
%
%
%
\subsection{Expression valid up to second order in $\Delta$}
To obtain the expression valid up to second order in $\Delta<<1$, we note  that $n^2=n_1^2(1-2\Delta f(r))$ and  use the expansion, $ln(1-x)=-x-x^2/2+...$ in Eq. 45 to arrive at
\begin{equation}
\delta\beta^2_{l\pm1}=\frac{2\pi\Delta}{N_l}\Big[\int_0^\infty(1+2\Delta f(r))\frac{\partial f(r)}{\partial r}F_{l}\big(\frac{dF_{l}(r)}{dr}\mp \frac{l}{r}F_{l}\big)rdr+\int_0^\infty\frac{\partial f(r)}{\partial r}G_{l\pm2}\big(\frac{dF_{l}(r)}{dr}\mp \frac{l}{r}F_{l}\big)rdr\Big].
\end{equation}
If we ignore the second term and replace $1+2\Delta f(r)$ by unity in Eq. 46, we obtain the result in \cite{snyder,bhandari2}, which is the correction to first order in $\Delta$. We consider now, for illustrative purposes, the step-index fiber.
\subsection{Application to a step-index fiber}
%
$n^2=n_1^2(1-2\Delta f(r))$, where  $\Delta <<1$. As a consequence, to first order in $\Delta$, $\partial ln(n^2)/\partial r=-2\Delta \partial f(r)/\partial r=-2\Delta \delta(r-a)$, the last step following from the fact that $f(r)$ is step function at $r=a$.
 Therefore, Eq. 46 reduces to 
\begin{equation}
\delta\beta^2_{l\pm1}=\frac{2\pi\Delta}{N_l}\Big[(1+2\Delta f(a))F_l(F'_l(a)\mp\frac{l}{a}F_l(a))a + G_{l+2}(a)((F'_l(a)\mp\frac{l}{a}F_l(a))a\Big],
\end{equation}
where $F'_l(a)$ is $\partial F_l(r)/\partial r$ evaluated at $r=a$. 
Furthermore, as noted in the previous section,
\begin{equation}
 F_{l}=(1/\sqrt{N_{l}}) J_l(\tilde{p}_{l}r),
\end{equation}
\begin{equation}
 G_{l\pm2}=\mu_{l\pm1}(1/\sqrt{N_{l}})J_{l\pm2}(\tilde{p}_{l}r),
\end{equation}
where  we have introduced a normalization constant, $N_l=2\pi\int_0^\infty F^2_l(r)rdr$, and $\mu_{l\pm 1}$ represents the relative amplitude of the second component given by the expression in Eqs. 33 and 38.
Substituting these expressions for $F_l(r)$ and $G_{l\pm2}(r)$ in Eq. 47, and setting $f(a)=1/2$, the (mean) value of the step function at $r=a$, the polarization correction to second order in $\Delta$ for the step-index fiber is found to be
\begin{equation}
\delta \beta_{l\pm1}^{2}=\mp(2\pi\Delta/N_{l})\big[(1+\Delta)\tilde{p}_{l}aJ_{l}(\tilde{p}_{l}a)J_{l\pm1}(\tilde{p}_{l}a)+\mu_{l\pm1}\tilde{p}_{l}aJ_{l\pm2}(\tilde{p}_{l}a)J_{l\pm1}(\tilde{p}_{l}a)\big].
\end{equation}
 $\mu_{l\pm1}=-(1/2)\Delta(\kappa_{l\pm1}/\alpha_{l\pm1}), \kappa_{l\pm 1}=-K_{l}(\tilde{q}_{l}a)/(\tilde{q}_{l}aK_{l\pm1}(\tilde{q}_{l}a))\mp(l+1)/(\tilde{q}_{l}a)^2$ and $\alpha_{l\pm1} = \pm J_{l}(\tilde{p}_{l}a)/(\tilde{p}_{l}aJ_{l\pm1}(\tilde{p}_{l}a)\mp(l\pm1)/(\tilde{p}_{l}a)^2)\pm\kappa_{l\pm1}$; we have used the fact that $p_{l\pm 1}\approx \tilde{p}_{l}$ and $q_{l\pm1}\approx \tilde{q}_{l}$ and the appropriate Bessel function identities. The first term on the RHS of Eq. 50 is the contribution of the first integral in Eq. 46 up to second order in $\Delta$ (without the $1+\Delta$ factor, it is the  correction to first order in $\Delta$ \cite{ bhandari2}). The second term in Eq. 50 is  of order $\Delta^2$ and  represents the impact of the polarization induced OAM field.  From Eq. 50, $\delta \beta_{l\pm1} = \delta \beta_{l\pm1}^{2}/(2\beta_{l\pm1})\approx \delta \beta_{l,\pm1}^{2}/(2\tilde{\beta}_l)$  follows. 
%
\subsection{Numerical results}
For numerical simulation we chose the multimode step-index fiber with parameters: $n_1=1.461, n_2=1.444, a=25 \mu m$, and $\lambda = 1.55 \mu m$. The index profile height  parameter $\Delta =0.011568$. We first calculated $\delta\beta_{l+1}^{(CE)}=\beta_{l+1}-\tilde{\beta_l}$ and $\delta\beta_{l-1}^{(CE)}=\beta_{l-1}-\tilde{\beta}_l$, where  $\beta_{l+1}$ and $\beta_{l-1}$ were determined  from the characteristic equations for the $HE_{l+1}$ and the $EH_{l-1}$ modes, respectively, while the  propagation constant, $\tilde{\beta}_l$ was determined  from the  characteristic equation for the scalar modes\cite{snyder}.  Values are displayed in Table 4 for $l=2$ to $l=5$.   $\delta\beta_{l\pm1}$, calculated directly from Eq. 50,  is in agreement with the characteristic equation-based values $\delta\beta_{l\pm1}^{(CE)}$ within a fraction of a percent.  The fifth column is the contribution of the polarization-induced OAM component, denoted $\delta\beta_{l\pm1}^{(p)}$, which is  calculated from the second term in Eq. 50. It  is  observed  to be about 1\% of the total, which is  of the order of $\Delta$, as expected.  The fractional differences (in column 4) are likely due to higher order terms not considered here 
%
All numerical calculations were done in MATLAB in double precision. 
%
\begin{table}
\centering
\caption{Polarization corrections, $\delta\beta_{l\pm1}^{(CE)}$ (in units of $meter^{-1}$) as calculated from the characteristic equations \cite{snyder}, and $\delta\beta_{l\pm1}$ as calculated directly from the analytic expression,  Eq. 50;  the  differences are of the order of a tenth of a percent; the fifth column is the contribution of the second term in Eq. 50.} 
\begin{tabular}{|c|c|c|c|c|} 
 \hline
$ l$&$\delta\beta_{l+1}^{(CE)}$&$\delta\beta_{l+1}$&\%$diff $&$\delta\beta_{l+1}^{(p)}$\\ 
\hline
2&-3.50859&-3.50404&0.13&0.03716\\ 
 \hline
 3&-5.59284 &-5.58343&0.17&0.05935\\ 
 \hline
4&-8.15965 &-8.14574&0.17&0.08669\\
\hline
5&-11.22993&-11.20679&0.21&0.11944\\
\hline
\end{tabular}
\end{table}
\begin{table}
\centering
\begin{tabular}{ |c|c|c|c|c|} 
 \hline
$ l$&$\delta\beta_{l-1}^{(CE)}$&$\delta\beta_{l-1}$&\%$diff $&$\delta\beta_{l-1}^{(p)}$\\ 
\hline
2&-2.91831&-2.91505&0.11&0.04907\\ 
 \hline
 3&-4.25156&-4.24659&0.12&0.06203\\ 
 \hline
4&-5.64576 &-5.63868&0.13&0.07642\\ 
\hline
5&-7.04443&-7.03481&0.14&0.08994\\
\hline
 \end{tabular}
\end{table}
\subsection{Extension to ring-core fiber}
A similar calculation can also  be carried  out  for the ring-core-fiber (see Fig. 1) once the function $G_{l\pm 2}(r)$ has been extracted from the expression, Eq. 43; the refractive index function here is given by  $n^2(r)=n_1^2(1-2\Delta f(r))$, where $f(r)=g(r-a)+h(r-b); g(r-a)= 1$ for $r<a$ and equal to zero for $r\ge a$, and $h(r-b)=1$ for $r\ge b$ and equal to zero for $r<b$. For $\partial f(r)\partial r$ needed in Eq. 46,  we note here that  $\partial h(r-b)/\partial r= \delta(r-b)$ and $ \partial g(r-a)/\partial r =-\delta (r-a)$. 
%
%
%
\section{Summary and discussion}
We have critically investigated the nature of the transverse OAM fields in a multilayered cylindrical waveguide. In particular, in a detailed study of the step-index fiber (two-layered fiber), we have identified the polarization-induced component of the vector modal  field as a modified scalar OAM mode, a hitherto unknown result. 
 The intensity peak of this nontraditional OAM mode occurs at a larger or smaller radial distance in the characteristic  donut intensity pattern, depending upon whether the spin is aligned with OAM or antialigned with OAM as in the transverse fields of the $HE_{l+1}$ or the $EH_{l-1}$ mode, respectively.  An extension to other multilayered fibers such as the ring-core fiber shows  the modified scalar field as a general feature of the polarization-induced component.  
The  amplitude of this modified scalar mode is of order $\Delta$, the index profile height parameter; thus larger the value of $\Delta$, larger this amplitude. 
We also have developed analytic expressions for the contribution of this component  to polarization correction, not done before. Specifically, we have applied these expressions to the case of a step-index fiber to illustrate their impact  numerically. Such expressions along with the insight they provide can be useful  in the  analysis and design of fibers.    Note also that, while the  treatment here pertains to the multilayered fibers, the  result of a polarization-induced component being a modified scalar modal field is a generic result that will manifest in any type of fiber (graded-index, or otherwise), simply because $\tilde{p}_l \ne \tilde{p}_{l\pm 2}$ in general. 
\\\\
{\large\bf   Appendix}
%
%
%
%
%
\appendix
%
%
\section{Reduction of the vector OAM modes to their scalar forms for the step-index fiber in weakly guiding approximation (WGA)}
We first note that setting $n_2=n_1$ and $\beta_l=kn_1$ in the characteristic equation for the vector modes (Eq. 22) yields
\begin{equation}
\frac{J_l'(p_la)}{p_laJ_l(p_la)}+\frac{K_l'(q_la)}{qaK_l(q_la)}=\pm |l|\Big[\frac{1}{q_l^2a^2}+\frac{1}{p_l^2a^2}\Big],
\end{equation}
where the + sign corresponds to the $EH_l$ mode and the - sign to the $HE_l$ mode.  Using the identities:  $J_l'(x)=-J_{l+1}(x)+(l/x)J_l(x)$ and  $K_l'(x)=-K_{l+1}(x)+(l/x)K_l(x)$ for  the case of the + sign, and the identities: $J_l'(x)=J_{l-1}(x)-(l/x)J_l(x)$ and  $K_l'(x)=-K_{l-1}(x)-(l/x)K_l(x)$ for the - sign, it can then be shown that the  characteristic equation for $EH_{l-1}$ mode and the characteristic equation for the $HE_{l+1}$ mode coalesce into a single characteristic equation given by 
\begin{equation}
\frac{J_l(\tilde{p}_la)}{\tilde{p}_laJ_{l-1}(\tilde{p}_la)}=-\frac{K_{l}(\tilde{q}_la)}{\tilde{q}_laK_{l-1}(\tilde{q}_{l-1}a)},
\end{equation}
where $\tilde{p}_l = p_{l+1}~(HE~modes)~=p_{l-1}~(EH~modes)$ in the limit $n_1 \rightarrow n_2$.
This  is a form  commonly found in literature  in the context of $LP$ modes \cite{snyder, buck, yariv}.
\\\\
We now note that the expression for $B_l/A_l$ given in Eq. 21 reduces, upon substitution of Eq. 51, to
\begin{equation}
\frac{B_l}{A_l}=\pm\frac{i\beta_l}{\omega \mu_0}.
\end{equation}
Eqs. 9 and 10 can be recast as
\begin{equation}
e_r^{(l)}=\frac{i\beta_l }{p_l^2}A_l\Big[p_lJ_l'(p_lr)+\frac{B_l}{A_l}(il)(\frac{\mu_0\omega}{\beta_l r})J_l(pr)\Big],
\end{equation}
\begin{equation}
e_\theta^{(l)}=\frac{i\beta_l}{p_l^2}A_l\Big[\frac{il}{r}A_lJ_l'(p_lr)-\frac{B_l}{A_l}\frac{\mu_0\omega p_l}{\beta_l }J_l'(p_lr)\Big].
\end{equation}
\underline{EH modes}
\\\\
Inserting Eq. 53 with the $+$ sign and invoking the identity: $J_l'(x)=-J_{l+1}(x)+\frac{l}{x}J_l(x)$, one arrives at
\begin{equation}
e_r^{(l)}=A'_lJ_{l+1}(p_lr),
\end{equation}
\begin{equation}
e_\theta^{(l)}=(-i)A'_lJ_{l+1}(p_lr),
\end{equation}
where the constant $A'_l$ is given by
\begin{equation}
A'_l=\Big(\frac{-i\beta_l}{p_l}\Big)A_l.
\end{equation}
Now $\beta_l/p_l >>1$ because $p_l=(k^2n_1^2-\beta_l^2)^{1/2}$ is very small  compared to $\beta_l \approx kn_1$. Consequently,  $e_z^{(l)}=A_lJ_{l}(p_lr)$
is very small compared to $e^{(l)}_r$ and $e^{(l)}_\theta$, and thus neglected generally. 
Note that $e^{(l)}_r$ and $e^{(l)}_\theta$ have the same magnitude but differ in phase by $\pi/2$.
Substituting Eqs. 55 and 56 in Eq. 6, we obtain
\begin{equation}
\vec{e}_t^{(l)}=\sqrt{2}A'J_{l+1}(pr)e^{i(l+1)\theta}\vec{\epsilon}_-.
\end{equation}
 Changing $l \rightarrow l-1$, 
\begin{equation}
\vec{e}_t^{(l-1)}=\sqrt{2}A'J_l(p_{l-1}r)e^{il\theta}\vec{\epsilon}_-,
\end{equation}
which can be rewritten as 
\begin{equation}
\vec{e}_t^{(l-1)}=\sqrt{2}A'_{l-1}O_l\vec{\epsilon}_-,
\end{equation}
where 
\begin{equation}
O_l=J_l( p_{l-1}r)e^{il\theta}=J_l(\tilde{p}_{l}r)e^{il\theta};
\end{equation}
$O_l$ is the amplitude of the scalar OAM mode  with topological charge $l$; it is the solution of the scalar wave equation (Eq. 2) to which the vector wave equation (Eq. 1) reduces in the limit $\Delta \rightarrow 0$; parameter $\tilde{p}_l^2=k^2n_1^2-\tilde{\beta}_l^2$. Eq. 61 signifies a field with a total angular momentum $J=L+S = l-1$, since the polarization $\vec{\epsilon}_-$ corresponds to a spin $S=-1$.  It is a case of \emph{spin-OAM nonalignment}.
\\\\
\underline{HE modes}
\\\\
Taking the negative sign in Eq. 53, and following the same steps as for the EH mode, but invoking the identity: $J_l'(x)=J_{l-1}(x)-\frac{l}{x}J_l(x)$, one arrives at 
\begin{equation}
e_r^{(l)}(r)=-A'_lJ_{l-1}(p_lr)
\end{equation}
and 
\begin{equation}
e_\theta^{(l)}(r)=-iA'_lJ_{l-1}(p_lr),
\end{equation}
with  $e_z^{(l)}<<e_r^{(l)}, e_\theta^{(l)}$ in magnitude. Inserting the above expressions in Eq. 6, we obtain
\begin{equation}
\vec{e}^{(l)}_t=-\sqrt{2}A'_lJ_{l-1}(pr)e^{i(l-1)\theta}\vec{\epsilon}_+,
\end{equation}
which implies further
\begin{equation}
\vec{e}^{(l+1)}_t=-\sqrt{2}A'_{l+1}J_l(p_{l+1}r)e^{il\theta}\vec{\epsilon}_+ =-\sqrt{2}A'_{l+1}J_l(\tilde{p}_{l}r)e^{il\theta} \vec{\epsilon}_+.
\end{equation}
Here the total angular momentum $J=L+S=l+1$ , i.e., we have a mode  with  \emph{ spin-OAM alignment}; $J_l(\tilde{p}_{l}r)e^{il\theta}$ is the scalar spatial OAM amplitude $O_l$.
\section {Derivation of the general expression for polarization correction}
The transverse vector field can be written  in the weakly guiding approximation (WGA)  as (see Eq. 4)
\begin{equation}
\vec{e}^{(l\pm1)}_t =\vec{\tilde e}^{(l)}_t + \delta \vec{e}^{(l\pm1)}_t, 
\end{equation}
where $\vec{\tilde e}^{(l)}_t $ is the corresponding scalar field and $\delta \vec{e}^{(l\pm1)}_t$, the polarization-induced field, 
 can be expressed as
\begin{equation} 
\delta\vec{e}_t^{(l\pm1)}=G_{l\pm2}(r)e^{i(l\pm2)\theta)}\vec{\epsilon}_\mp.
\end{equation}
Substituting Eq. 67 in Eq. 1  and invoking Eq. 2, we obtain 
\begin{equation}
(\vec{\nabla}_t^2+k^2n^2)\delta \vec{e}^{(l\pm1)}_t +\vec{\nabla}_t(\vec{e}^{(l\pm1)}_t.\vec{\nabla}_t ln~ n^2)=\beta^2_{l\pm 1}\delta\vec{e}^{(l\pm 1)}_t +\delta\beta^2_{l\pm 1}\vec{\tilde{e}}_t^{(l)},
\end{equation}
where  $\delta\beta^2_{l\pm 1}=\beta^2_{l\pm 1}-\tilde{\beta}^2_l$ (see Eq. 3).  
Now  $G_{l\pm2}(r)=G_{l\pm2}(p_{l\pm1}r)$, where the parameter $p_{l\pm1}$ is given by $p^2_{l\pm1}=k^2n_1^2-\beta^2_{l\pm1}$. Furthermore, the function $G_{l\pm2}(p_{l\pm1}r)$ is the Bessel functions $J_{l\pm 2}(p_{l\pm 1}r)$ for the step index fiber and a linear combination of $J_{l\pm2}(p_{l\pm1}r)$ and $Y_{l\pm2}(p_{l\pm1}r)$ for the ring-core fiber. Noting these facts it then follows from a direct substitution of Eq. 68 that
\begin{equation}
(\vec{\nabla}_t^2+k^2n^2)\delta \vec{e}^{(l\pm 1)}_t=(k^2n_1^2-p^2_{l\pm 1})\delta \vec{e}^{(l+1)}_t=\beta_{l\pm 1}^2\delta \vec{e}^{(l\pm 1)}_t.
\end{equation}
Inserting Eq. 70 into Eq. 69 yields
\begin{equation}
\delta\beta^2_{l\pm 1}\vec{\tilde{e}}_t^{(l)}=\vec{\nabla}_t(\vec{e}^{(l\pm 1)}_t.\vec{\nabla}_t ln~ n^2)
\end{equation}
leading to
\begin{equation}
\delta\beta_{l\pm 1}^2 =<\vec{\tilde{e}}_t^{(l)} |\vec{\nabla}_t^{(l)}(\vec{e}^{(l\pm 1)}_t.\vec{\nabla}_t ln~ n^2)>=\int_0^{\infty}\int_0^{2\pi}\vec{\tilde{e}}^{(l)*}_t.\vec{\nabla}_t(\vec{e}^{(l\pm 1)}_t.\vec{\nabla}_t ln~ n^2)r dr d\theta.
\end{equation}
Here we have assumed $\vec{\tilde{e}}_t^{(l)} $ is normalized, i.e., $<\vec{\tilde{e}}_t^{(l)} |\vec{\tilde{e}}_t^{(l)} >=1$; the bra ($<$) and  ket ($>$) notation indicates a scalar product. Substitution of $\vec{e}^{(l\pm1)}_t =\vec{\tilde e}^{(l)}_t+\delta \vec{e}^{(l\pm 1)}_t$ in Eq. 72 splits the  integral into two integrals: 
\begin{equation}
\delta\beta_{l\pm 1}^2 =\int_0^{\infty}\int_0^{2\pi}\vec{\tilde{e}}^{(l)*}_t.\vec{\nabla}_t(\vec{\tilde{e}}^{(l)}_t.\vec{\nabla}_t ln~ n^2)r dr d\theta +\int_0^{\infty}\int_0^{2\pi}\vec{\tilde{e}}^{(l)*}_t.\vec{\nabla}_t(\delta \vec{e}^{(l\pm 1)}_t.\vec{\nabla}_t ln~ n^2)r dr d\theta.
\end{equation}
Substituting further $\vec{\tilde{e}}_t=F_le^{il\theta}\vec{\epsilon}_\pm$ and 
$\delta\vec{e}^{(l\pm 1)}_t=G_{l\pm2}(r)e^{i(l\pm 2)\theta)}\vec{\epsilon}_\mp$ (Eq. 68),  performing integration by parts, and noting that $F_l(r)$ and $G_{l\pm 2} (r)$ vanish at the end points of the integral, $r=0$ and $r=\infty$, we arrive at Eq. 45:
\begin{equation}
\delta\beta^2_{l,\pm1}=-\pi\Big[\int_0^\infty\frac{\partial ln(n^2)}{\partial r}F_{l}(r)\Big(\frac{dF_{l}(r)}{dr}\mp \frac{l}{r}F_{l}\Big)rdr+\int_0^\infty\frac{\partial ln(n^2)}{\partial r}G_{l\pm2}\Big(\frac{dF_{l}(r)}{dr}\mp \frac{l}{r}F_{l}(r)\Big)rdr\Big];
\end{equation}
we have assumed, as before, $n^2=n^2(r)$ and also used the fact that $\vec{\epsilon}_{\pm}=1/\sqrt{2}(\hat{x}\pm i\hat{y})=1/\sqrt{2}(\hat{r}\pm i\hat{\theta})e^{\pm i\theta}$ (in polar coordinates).
%
%
%
%

%
\end{document}